\documentclass[prl,twocolumn,showpacs]{revtex4}
\usepackage{amsfonts,amssymb,amsmath,bm,graphics}

\newcommand{\f}{\frac}
\newcommand{\p}{\partial}
\newcommand{\rmT}{\mathrm{T}}
\newcommand{\deb}{\mathrm{De}}
\newcommand{\rey}{\mathrm{Re}}
\newcommand{\bfT}{\mathbf{T}}
\newcommand{\bme}{\bm{e}}
\newcommand{\bmu}{\bm{u}}
\newcommand{\bcdot}{\bm{\cdot}}
\newcommand{\bnabla}{\bm{\nabla}}

\begin{document}

\title{Magnetorotational-type Instability in Couette--Taylor Flow\\
of a Viscoelastic Polymer Liquid}

\author{Gordon I. Ogilvie}
\email{gio10@cam.ac.uk}
\author{Adrian T. Potter}
\affiliation{Department of Applied Mathematics and Theoretical Physics,
University of Cambridge, Centre for Mathematical Sciences, Wilberforce Road,
Cambridge CB3 0WA, UK}

\date{\today}

\begin{abstract}
We describe an instability of viscoelastic Couette--Taylor flow that
is directly analogous to the magnetorotational instability (MRI) in
astrophysical magnetohydrodynamics, with polymer molecules playing the
role of magnetic field lines.  By determining the conditions required
for the onset of instability and the properties of the preferred
modes, we distinguish it from the centrifugal and elastic
instabilities studied previously.  Experimental demonstration and
investigation should be much easier for the viscoelastic instability
than for the MRI in a liquid metal.  The analogy holds with the case
of a predominantly toroidal magnetic field such as is expected in an
accretion disk and it may be possible to access a turbulent regime in
which many modes are unstable.
\end{abstract}

\pacs{47.20.-k, 46.35.+z, 47.65.-d, 97.10.Gz}

\maketitle

The magnetorotational instability (MRI) is widely regarded as one of
the most important instabilities in astrophysics.  It occurs in a
differentially rotating fluid of sufficiently high electrical
conductivity in the presence of a magnetic field, when the angular
velocity decreases with distance from the axis of rotation.  The MRI
is especially relevant to astrophysical accretion disks, where it is
believed to give rise to turbulence and angular momentum transport,
but it can also occur in stellar interiors \cite[and references
therein]{BH98,B03}.

The mechanism of instability can be visualized by considering two
fluid elements connected by magnetic field lines, which act as an
elastic tether.  The elements initially orbit at the same radius and
one is displaced inward so that its angular velocity increases.
Being retarded by the tether, it loses angular momentum and moves to a
still smaller orbit so that the process runs away.  Both inertial
forces and high conductivity are important to the mechanism.

Experimental demonstration of the MRI is regarded as highly desirable
but has proved difficult because of the need to achieve a sufficiently
large magnetic Reynolds number ($\mathrm{Rm}=LU\mu_0\sigma$, where $L$
and $U$ are characteristic length and velocity scales of the flow and
$\sigma$ is the electrical conductivity).  Even for the most highly
conducting liquid metals, the required Reynolds number ($\rey=LU/\nu$,
where $\nu$ is the kinematic viscosity) is extremely large, in excess
of $10^6$.

A natural choice of experiment is the Couette--Taylor flow of a liquid
metal between differentially rotating concentric cylinders.  By
setting the angular velocities of the cylinders in a ``Keplerian''
ratio, $\Omega\propto r^{-3/2}$, and introducing a uniform axial
magnetic field, the MRI becomes possible while the centrifugal
(inertial) instability is suppressed according to Rayleigh's criterion
\cite{C61}.  The water prototype \cite{JBSG06} of a planned experiment
using liquid gallium \cite{LGJ06} has attempted to deal with the
strong end-effects that arise at large $\mathrm{Re}$ when the
cylinders are not much taller than the gap between them.

Behavior related to the MRI may have already been observed in a
\emph{spherical} Couette flow of liquid sodium \cite{S04}.  The
interpretation of this experiment is difficult because of the presence
of hydrodynamic turbulence and a complicated pattern of differential
rotation in the unmagnetized state.  A different instability, which
relies on the presence of a helical magnetic field, has also been
reported in a cylindrical experiment \cite{HR05,S06}.  This result is
controversial \cite{LGHJ06} and, since the instability occurs at
$\mathrm{Rm}\ll1$, it does not resemble the MRI as conventionally
understood.

In this Letter we describe an instability that is directly analogous
to the MRI and occurs instead in a simple viscoelastic liquid, with
polymer molecules playing the role of magnetic field lines.  We
suggest that an experiment to study this MRI analog would be much
easier and may in some respects represent the astrophysical systems of
interest more closely than liquid metal experiments.

The characteristic properties of magnetohydrodynamics (MHD) in the
limit of large $\mathrm{Rm}$ are that the magnetic field lines are
advected and distorted by the fluid flow and can be identified with
material lines, and that the field exerts a tension through the
Lorentz force.  These two properties imply that the magnetic field
imparts an elasticity to an electrically conducting fluid, which
allows the propagation of Alfv\'en waves, for example.

Directly analogous behavior occurs in viscoelastic fluids such as
dilute solutions of long-chain flexible polymer molecules in water or
organic solvents.  The polymer molecules also impart an elasticity to
the fluid because they are advected and distorted by the flow and
exert a tension.  In earlier work \cite{OP03} we drew attention to the
physical and mathematical similarities between these situations.  The
mathematical viewpoint is that the force per unit volume is the
divergence of a symmetric stress tensor which is Lie-transported by
the fluid flow in the sense of a second-rank contravariant tensor
field (its upper-convected derivative is zero).  This result applies
asymptotically both to the (Maxwell) magnetic stress in MHD in the
limit of large $\mathrm{Rm}$ and to the polymeric stress in the limit
of large Deborah number ($\deb=S\tau$, where $S$ is the shear rate and
$\tau$ is the relaxation time of the fluid).

We also showed \cite{OP03} that there is a direct analog of the MRI in
rotating shear flows of a viscoelastic fluid, using a Cartesian model
that can be considered as a local approximation of a Couette--Taylor
flow when the rotation and shear rates are comparable.  For $\deb\gg1$
the polymeric tension is parallel to the flow and corresponds to an
equivalent magnetic field also in that (azimuthal) direction.  Growth
rates and eigenfunctions of the viscoelastic and MHD problems were
found to be very similar.

Here we present a new analysis of the linear stability of viscoelastic
Couette--Taylor flow, applicable to an experiment with cylindrical
geometry.  By optimizing the growth rate over all wavenumbers, we
determine the likely initial outcome of such an experiment.  We also
show how the MRI analog can be distinguished from other types of
instability in the same system.

A viscoelastic fluid is governed by the equation of motion,
\begin{equation}
  \rho\left(\f{\p\bmu}{\p t}+\bmu\bcdot\bnabla\bmu\right)=-\bnabla p+
  \bnabla\bcdot\bfT+\mu_s\nabla^2\bmu,
\label{motion}
\end{equation}
the incompressibility condition, $\bnabla\bcdot\bmu=0$, and, in the
case of an Oldroyd-B fluid, the constitutive equation
\begin{equation}
  \begin{split}
    &\bfT+\tau\left[\f{\p\bfT}{\p t}+\bmu\bcdot\bnabla\bfT-
    (\bnabla\bmu)^\rmT\bcdot\bfT-\bfT\bcdot\bnabla\bmu\right]\\
    &\qquad=\mu_p\left[\bnabla\bmu+(\bnabla\bmu)^\rmT\right].
  \end{split}
\label{constitutive}
\end{equation}
Here $\rho$, $p$, and $\bmu$ are the density, pressure, and velocity of
the fluid, $\mu_s$ and $\mu_p$ are the solvent and polymer
viscosities, and $\tau$ is the relaxation time of the polymeric stress
$\bfT$.  The widely adopted Oldroyd-B model is the simplest nonlinear
description of a dilute polymer solution and can be derived directly
from a kinetic model \cite{BCAO87}.

In the Couette--Taylor experiment the basic flow is
$\bmu=r\Omega(r)\,\bme_\phi$, where $(r,\phi,z)$ are cylindrical polar
coordinates, $\Omega(r)=A+Br^{-2}$, and the constants $A$ and $B$ are
such as to match the angular velocities $\Omega_{1,2}$ of the inner
and outer cylinders of radii $r_{1,2}$.  The polymeric stress
components are $T_{r\phi}=-2\mu_pBr^{-2}$ and
$T_{\phi\phi}=8\mu_p\tau B^2r^{-4}$, the latter being analogous to
the ``hoop stress'' of a magnetic field
$B_\phi=(\mu_0T_{\phi\phi})^{1/2}$.  End-effects should be unimportant
provided that the height of the cylinders is much larger than the gap
width $d=r_2-r_1$.

Two dimensionless numbers associated with the experiment are the
Reynolds number $\rey=r_1d|\Omega_2-\Omega_1|/\nu$, which measures the
relative importance of inertial and viscous effects
(here $\nu=(\mu_s+\mu_p)/\rho$ is the total kinematic viscosity),
and the Deborah number $\deb=r_1|\Omega_2-\Omega_1|\tau/d$, which
quantifies the importance of elasticity.

We linearize equations~(\ref{motion}) and~(\ref{constitutive}) about
the basic flow and seek solutions with the dependence
$\exp(st+im\phi+ikz)$, obtaining a sixth-order system of ordinary
differential equations for the radial structure of an eigenmode with
complex growth rate $s$, integer azimuthal mode number $m$ and real
axial wavenumber $k$.  No-slip boundary conditions apply at
$r=r_{1,2}$.

We solve the problem by two independent methods.  First, we apply
Chebyshev collocation on a Gauss--Lobatto grid to convert the system
into an algebraic generalized eigenvalue problem, which we solve
directly using a LAPACK routine.  Second, we use a shooting method
based on a 4th/5th order Runge--Kutta integrator with adaptive
stepsize.  The two routines agree to very high precision.  The
Chebyshev method is useful for examining all the eigenmodes with a
sufficiently simple radial structure to be adequately resolved
(spurious modes are also generated which must be rejected).  The
shooting method is good for following a single mode as the parameters
are varied.

At each point in the $(\deb,\rey)$ plane we seek to maximize the
growth rate $\mathrm{Re}(s)$ with respect to both $m$ and $k$.  Once
the preferred wavenumbers are obtained at one point, the optimal
solution is tracked through the plane, making the necessary
adjustments to $m$ and $k$.

Two different instabilities have previously been reported in
viscoelastic Couette--Taylor flow \cite{L92}.  One is the centrifugal
(inertial) instability (CI) of Rayleigh and Taylor \cite{C61}, which
occurs at sufficiently large $\rey$ even in the Newtonian limit
($\deb=0$).  The other is the purely elastic instability (EI) of
Muller at~al.~\cite{MLS89} and Larson et~al.~\cite{LSM90}, which
occurs at sufficiently large $\deb$ even in the inertialess limit
($\rey=0$).  The CI is driven by shear energy and requires the
specific angular momentum $|r^2\Omega|$ to decrease with $r$, while
the EI is driven by elastic energy and simply requires shear and
curved streamlines.

The MRI analog, however, depends on both elasticity and inertial
forces, and therefore requires both $\rey$ and $\deb$ to be
sufficiently large.  It is driven by shear energy and requires the
angular velocity $|\Omega|$ to decrease with $r$ \cite{C61}.

We tested our methods by reproducing standard results on the CI of a
Newtonian fluid and also the results of Ref.~\cite{OP03} in the limit
of a narrow gap.

We made strenuous efforts to reproduce the results of
Ref.~\cite{AB93}, where a problem identical to ours was considered in
a different parameter regime, but we obtained only qualitative
agreement with many of their findings.  To verify our own results we
derived the linearized equations by two independent methods and also
obtained matching numerical results with completely independent codes.

For the purposes of illustration we adopt the parameters
$r_1/r_2=0.95$ and $\mu_s/\mu_p=1$.  The latter ratio could be
either small or large in practice, depending on the choice of solvent.
Boger fluids, which avoid shear thinning and conform more accurately
to the Oldroyd-B model, have more viscous solvents.

We set the angular velocities of the cylinders in the ratio
$\Omega_2/\Omega_1=(r_2/r_1)^p$, with $p=-3/2$ (case~1), $p=+3/2$
(case~2) or $p=-3$ (case~3).  Case~1 (a Keplerian ratio) should
exhibit both EI and MRI, case~2 only EI, and case~3 all three
instabilities.

In Fig.~1 we show the optimized growth rate in the $(\deb,\rey)$ plane
for case~1.  The EI is seen at $\rey=0$ for $\deb>19$; it prefers a
small value of $m$ and its growth rate is limited here to
$<0.1\,\Omega_1$.  As $\rey$ is increased, instability occurs for a
much wider range of $\deb$ and with larger growth rates.  This is the
MRI analog.  The preferred value of $m$ increases with $\rey/\deb$
as the equivalent magnetic field becomes weaker \cite{OP03}.  The EI
prefers a larger value of $k$.
As expected, the CI is absent and there is no instability in the
Newtonian limit $\deb=0$.

\begin{figure}
  \includegraphics{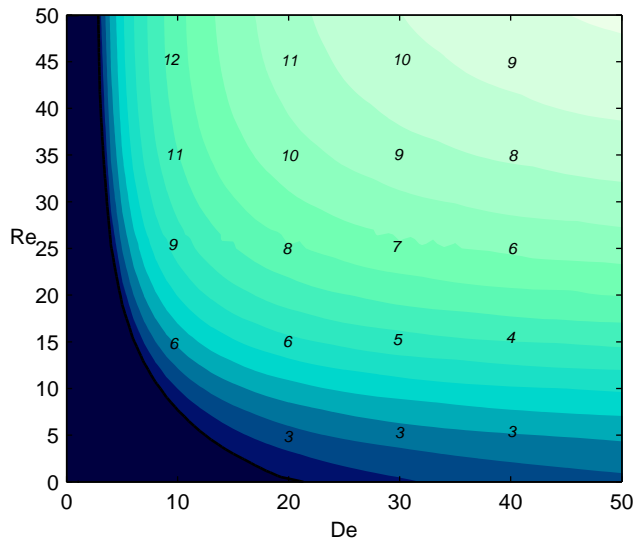}
  \caption{Contours of the growth rate $\mathrm{Re}(s)/\Omega_1$,
  optimized with respect to~$m$ and~$k$, for case~1 ($p=-3/2$).
  Contour values increase from~$0$ to~$0.35$ in steps of $0.025$
  starting from the thick line (negative contours are not plotted).
  The preferred azimuthal mode number $m$ is indicated by numerals.}
\end{figure}

A typical eigenfunction of the MRI analog at the onset of instability
is shown in Fig.~2.  This solution is qualitatively similar to Taylor
vortices and has a vertical wavelength comparable to the gap width but
is nonaxisymmetric and would appear as a rotating spiral pattern.
However, nonlinear pattern selection may favor a linear combination of
upward and downward spirals, resulting in a ``ribbon'' structure.

\begin{figure}
  \includegraphics{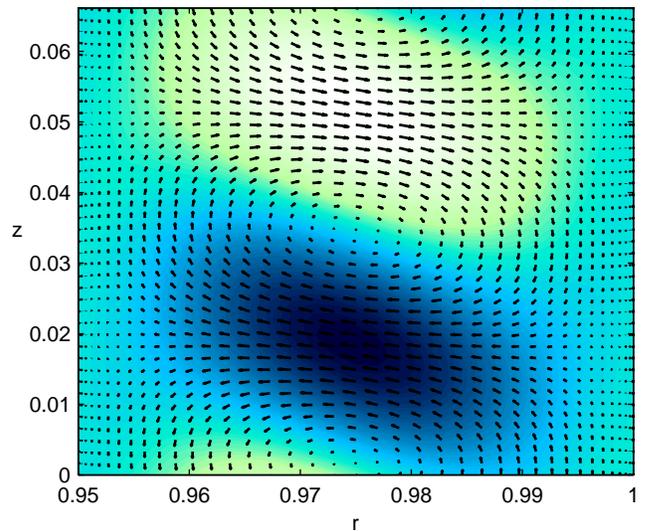}
  \caption{Eigenfunction of the preferred mode in case~1 at the onset
  of instability for $(\deb,\rey)=(3.3,40)$, where $(m,k)=(11,95)$.
  The azimuthal velocity perturbation is indicated by
  color/grayscale.  One vertical wavelength is shown.}
\end{figure}

In case~2 (Fig.~3) the ratio of shear to rotation is similar to case~1
but of opposite sign, so that the MRI and CI are absent.  We find that
the EI operates in a very similar fashion at small $\rey$ but is
suppressed by increasing $\rey$ at fixed $\deb$.  A comparison of
Figs~1 and~3 makes it clear that only the very low-$\mathrm{Re}$
behavior in Fig.~1 can be attributed to the EI.

\begin{figure}
  \includegraphics{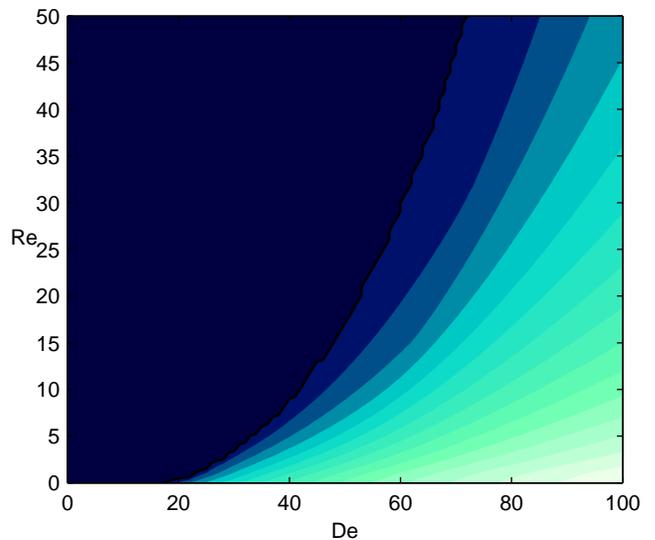}
  \caption{Growth rate for case~2 ($p=+3/2$).  Contour values increase
  from~$0$ to~$0.07$ in steps of $0.005$.  The preferred value of~$m$
  is~$2$ throughout.  Note the different horizontal scale.}
\end{figure}

In case~3 (Fig.~4) we observe similar behavior to case~1 in much of
the $(\deb,\rey)$ plane.  However, the CI is seen operating at small
$\deb$ for $\rey>85$, where it prefers $m=0$ (the classic axisymmetric
Taylor vortices).  Fig.~4 therefore illustrates EI, MRI, and CI, but
most of the unstable region can be identified as MRI by comparison
with Figs~1 and~3.  The growth rates and preferred mode numbers of the
MRI are somewhat different from case~1 because of the different ratio
of shear to rotation.  We find that the three instabilities are
described by a mode that changes its character in a continuous way
across the parameter space.

\begin{figure}
  \includegraphics{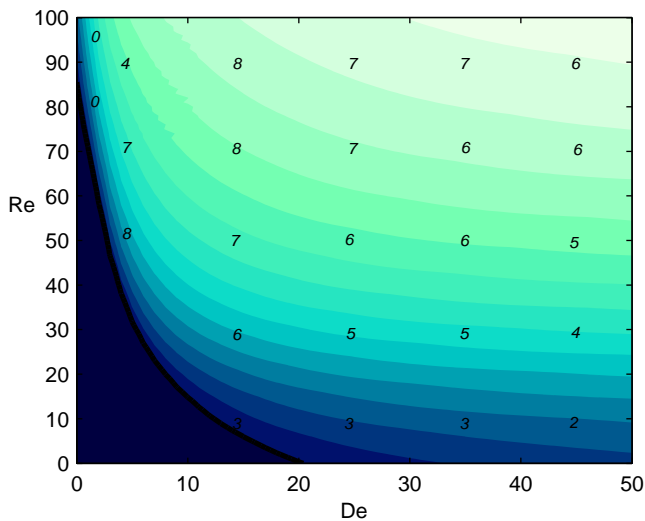}
  \caption{Growth rate for case~3 ($p=-3$).  Contour values increase
  from~$0$ to~$0.7$ in steps of $0.05$.  Note the different vertical
  scale.}
\end{figure}

Experimental and theoretical studies of instabilities in viscoelastic
Couette--Taylor flow have been carried out since the 1960s but have
not revealed the three types of characteristic behavior seen here.
The MRI analog is exhibited most clearly, as in our case~1, when the
angular velocity decreases outward while the angular momentum
increases.  The few experiments in which the inner and outer cylinders
can rotate independently (e.g.,~\cite{BM97}) appear not to have
examined this regime.

Previous studies have described either a reduction of the critical
$\rey$ for the CI as $\deb$ is increased \cite[and references
therein]{L92} or a reduction of the critical $\deb$ for the EI as
$\rey$ is increased \cite{GS98}.  The physical nature of this intermediate
``inertioelastic mode'' has not been discussed.  A similar behavior
can be seen in case~3 (Fig.~4) but we interpret it as the MRI analog.
The MRI tends to dominate the CI when both the angular velocity and
the angular momentum decrease outward, unless $\deb$ is small.
In case~2 (Fig.~3), where the MRI is absent, increasing $\rey$ tends
instead to suppress the EI.

An experimental study of the MRI analog would require a standard
Couette--Taylor apparatus and a suitable polymer solution.  The
polymer viscosity and relaxation time can be adjusted by varying the
concentration of the solution.  Care must be taken to maintain the
temperature of the fluid in the presence of viscous dissipation.  Some
effects such as shear-thinning in lower-viscosity solvents may require
the use of a more sophisticated constitutive model than Oldroyd-B.
Subcritical, hysteretic behavior has been reported in viscoelastic
Couette--Taylor flow \cite{GS98,TAS06} and this may also affect the
experimental manifestation of the MRI analog.

The asymptotic nature of the analogy between viscoelastic and MHD
flows \cite{OP03} means that it is more accurate for larger values of
$\mathrm{Re}$ and $\mathrm{De}$.  At the onset of instability, as in
Fig.~2, the analogy is good but not perfect.  It is better in the
upper right part of Fig.~1, where many modes are unstable and the flow
may become turbulent.  This situation would be analogous to a
turbulent accretion disk containing a predominantly toroidal magnetic
field.  In contrast, liquid metal MRI experiments employ an axial
magnetic field and are not expected to be able to access a regime of
MHD turbulence.  In these respects the polymer experiment may offer a
more faithful representation of the astrophysical systems of interest.
Other major advantages are that the materials are cheap and safe, and
that optical visualization or velocimetry can be carried out.

\begin{acknowledgments}
This work was supported in part by STFC.

\end{acknowledgments}

\end{document}